\documentclass[conference,10pt,twocolumn,a4paper]{IEEEtran}

\usepackage{cite}
\usepackage[T1]{fontenc}
\usepackage{graphicx}
\usepackage{amssymb}
\usepackage{amsmath}
\usepackage{paralist}
\usepackage{setspace}
\usepackage{microtype}
\usepackage{hyperref}
\usepackage{url}
\usepackage{balance}
\usepackage[caption=false,font=footnotesize]{subfig}
\usepackage{xcolor}
\usepackage{fancyref}
\usepackage{dsfont}

\IEEEoverridecommandlockouts

\usepackage{amssymb}
\usepackage{amsfonts}
\usepackage{mathrsfs}
\usepackage{xspace}
\usepackage{bm}
\usepackage{upgreek}

\newcommand{\safemath}[2]{\newcommand{#1}{\ensuremath{#2}\xspace}}

\safemath{\bma}{\mathbf{a}}
\safemath{\bmb}{\mathbf{b}}
\safemath{\bmc}{\mathbf{c}}
\safemath{\bmd}{\mathbf{d}}
\safemath{\bme}{\mathbf{e}}
\safemath{\bmf}{\mathbf{f}}
\safemath{\bmg}{\mathbf{g}}
\safemath{\bmh}{\mathbf{h}}
\safemath{\bmi}{\mathbf{i}}
\safemath{\bmj}{\mathbf{j}}
\safemath{\bmk}{\mathbf{k}}
\safemath{\bml}{\mathbf{l}}
\safemath{\bmm}{\mathbf{m}}
\safemath{\bmn}{\mathbf{n}}
\safemath{\bmo}{\mathbf{o}}
\safemath{\bmp}{\mathbf{p}}
\safemath{\bmq}{\mathbf{q}}
\safemath{\bmr}{\mathbf{r}}
\safemath{\bms}{\mathbf{s}}
\safemath{\bmt}{\mathbf{t}}
\safemath{\bmu}{\mathbf{u}}
\safemath{\bmv}{\mathbf{v}}
\safemath{\bmw}{\mathbf{w}}
\safemath{\bmx}{\mathbf{x}}
\safemath{\bmy}{\mathbf{y}}
\safemath{\bmz}{\mathbf{z}}
\safemath{\bmzero}{\mathbf{0}}
\safemath{\bmone}{\mathbf{1}}

\bmdefine{\biad}{a}
\bmdefine{\bibd}{b}
\bmdefine{\bicd}{c}
\bmdefine{\bidd}{d}
\bmdefine{\bied}{e}
\bmdefine{\bifd}{f}
\bmdefine{\bigd}{g}
\bmdefine{\bihd}{h}
\bmdefine{\biid}{i}
\bmdefine{\bijd}{j}
\bmdefine{\bikd}{k}
\bmdefine{\bild}{l}
\bmdefine{\bimd}{m}
\bmdefine{\bind}{n}
\bmdefine{\biod}{o}
\bmdefine{\bipd}{p}
\bmdefine{\biqd}{q}
\bmdefine{\bird}{r}
\bmdefine{\bisd}{s}
\bmdefine{\bitd}{t}
\bmdefine{\biud}{u}
\bmdefine{\bivd}{v}
\bmdefine{\biwd}{w}
\bmdefine{\bixd}{x}
\bmdefine{\biyd}{y}
\bmdefine{\bizd}{z}

\bmdefine{\bixid}{\xi}
\bmdefine{\bilambdad}{\lambda}
\bmdefine{\bimud}{\mu}
\bmdefine{\bithetad}{\theta}
\bmdefine{\biphid}{\phi}
\bmdefine{\bideltad}{\delta}

\safemath{\bmia}{\biad}
\safemath{\bmib}{\bibd}
\safemath{\bmic}{\bicd}
\safemath{\bmid}{\bidd}
\safemath{\bmie}{\bied}
\safemath{\bmif}{\bifd}
\safemath{\bmig}{\bigd}
\safemath{\bmih}{\bihd}
\safemath{\bmii}{\biid}
\safemath{\bmij}{\bijd}
\safemath{\bmik}{\bikd}
\safemath{\bmil}{\bild}
\safemath{\bmim}{\bimd}
\safemath{\bmin}{\bind}
\safemath{\bmio}{\biod}
\safemath{\bmip}{\bipd}
\safemath{\bmiq}{\biqd}
\safemath{\bmir}{\bird}
\safemath{\bmis}{\bisd}
\safemath{\bmit}{\bitd}
\safemath{\bmiu}{\biud}
\safemath{\bmiv}{\bivd}
\safemath{\bmiw}{\biwd}
\safemath{\bmix}{\bixd}
\safemath{\bmiy}{\biyd}
\safemath{\bmiz}{\bizd}

\safemath{\bmxi}{\bixid}
\safemath{\bmlambda}{\bilambdad}
\safemath{\bmmu}{\bimud}
\safemath{\bmtheta}{\bithetad}
\safemath{\bmphi}{\biphid}
\safemath{\bmdelta}{\bideltad}

\safemath{\bA}{\mathbf{A}}
\safemath{\bB}{\mathbf{B}}
\safemath{\bC}{\mathbf{C}}
\safemath{\bD}{\mathbf{D}}
\safemath{\bE}{\mathbf{E}}
\safemath{\bF}{\mathbf{F}}
\safemath{\bG}{\mathbf{G}}
\safemath{\bH}{\mathbf{H}}
\safemath{\bI}{\mathbf{I}}
\safemath{\bJ}{\mathbf{J}}
\safemath{\bK}{\mathbf{K}}
\safemath{\bL}{\mathbf{L}}
\safemath{\bM}{\mathbf{M}}
\safemath{\bN}{\mathbf{N}}
\safemath{\bO}{\mathbf{O}}
\safemath{\bP}{\mathbf{P}}
\safemath{\bQ}{\mathbf{Q}}
\safemath{\bR}{\mathbf{R}}
\safemath{\bS}{\mathbf{S}}
\safemath{\bT}{\mathbf{T}}
\safemath{\bU}{\mathbf{U}}
\safemath{\bV}{\mathbf{V}}
\safemath{\bW}{\mathbf{W}}
\safemath{\bX}{\mathbf{X}}
\safemath{\bY}{\mathbf{Y}}
\safemath{\bZ}{\mathbf{Z}}

\safemath{\bZero}{\mathbf{0}}
\safemath{\bOne}{\mathbf{1}}
\safemath{\bDelta}{\mathbf{\Delta}}
\safemath{\bLambda}{\mathbf{\UpLambda}}
\safemath{\bPhi}{\mathbf{\Upphi}}
\safemath{\bSigma}{\mathbf{\Upsigma}}
\safemath{\bOmega}{\mathbf{\Upomega}}
\safemath{\bTheta}{\mathbf{\Uptheta}}

\bmdefine{\biAd}{A}
\bmdefine{\biBd}{B}
\bmdefine{\biCd}{C}
\bmdefine{\biDd}{D}
\bmdefine{\biEd}{E}
\bmdefine{\biFd}{F}
\bmdefine{\biGd}{G}
\bmdefine{\biHd}{H}
\bmdefine{\biId}{I}
\bmdefine{\biJd}{J}
\bmdefine{\biKd}{K}
\bmdefine{\biLd}{L}
\bmdefine{\biMd}{M}
\bmdefine{\biOd}{N}
\bmdefine{\biPd}{O}
\bmdefine{\biQd}{P}
\bmdefine{\biRd}{R}
\bmdefine{\biSd}{S}
\bmdefine{\biTd}{T}
\bmdefine{\biUd}{U}
\bmdefine{\biVd}{V}
\bmdefine{\biWd}{W}
\bmdefine{\biXd}{X}
\bmdefine{\biYd}{Y}
\bmdefine{\biZd}{Z}

\bmdefine{\biDelta}{\Delta}
\bmdefine{\biLambda}{\Lambda}
\bmdefine{\biPhi}{\Phi}
\bmdefine{\biSigma}{\Sigma}
\bmdefine{\biOmega}{\Omega}
\bmdefine{\biTheta}{\Theta}

\safemath{\bimA}{\biAd}
\safemath{\bimB}{\biBd}
\safemath{\bimC}{\biCd}
\safemath{\bimD}{\biDd}
\safemath{\bimE}{\biEd}
\safemath{\bimF}{\biFd}
\safemath{\bimG}{\biGd}
\safemath{\bimH}{\biHd}
\safemath{\bimI}{\biId}
\safemath{\bimJ}{\biJd}
\safemath{\bimK}{\biKd}
\safemath{\bimL}{\biLd}
\safemath{\bimM}{\biMd}
\safemath{\bimN}{\biNd}
\safemath{\bimO}{\biOd}
\safemath{\bimP}{\biPd}
\safemath{\bimQ}{\biQd}
\safemath{\bimR}{\biRd}
\safemath{\bimS}{\biSd}
\safemath{\bimT}{\biTd}
\safemath{\bimU}{\biUd}
\safemath{\bimV}{\biVd}
\safemath{\bimW}{\biWd}
\safemath{\bimX}{\biXd}
\safemath{\bimY}{\biYd}
\safemath{\bimZ}{\biZd}

\safemath{\bimDelta}{\biDelta}
\safemath{\bimLambda}{\biLambda}
\safemath{\bimPhi}{\biPhi}
\safemath{\bimSigma}{\biSigma}
\safemath{\bimOmega}{\biOmega}
\safemath{\bimTheta}{\biTheta}

\safemath{\setA}{\mathcal{A}}
\safemath{\setB}{\mathcal{B}}
\safemath{\setC}{\mathcal{C}}
\safemath{\setD}{\mathcal{D}}
\safemath{\setE}{\mathcal{E}}
\safemath{\setF}{\mathcal{F}}
\safemath{\setG}{\mathcal{G}}
\safemath{\setH}{\mathcal{H}}
\safemath{\setI}{\mathcal{I}}
\safemath{\setJ}{\mathcal{J}}
\safemath{\setK}{\mathcal{K}}
\safemath{\setL}{\mathcal{L}}
\safemath{\setM}{\mathcal{M}}
\safemath{\setN}{\mathcal{N}}
\safemath{\setO}{\mathcal{O}}
\safemath{\setP}{\mathcal{P}}
\safemath{\setQ}{\mathcal{Q}}
\safemath{\setR}{\mathcal{R}}
\safemath{\setS}{\mathcal{S}}
\safemath{\setT}{\mathcal{T}}
\safemath{\setU}{\mathcal{U}}
\safemath{\setV}{\mathcal{V}}
\safemath{\setW}{\mathcal{W}}
\safemath{\setX}{\mathcal{X}}
\safemath{\setY}{\mathcal{Y}}
\safemath{\setZ}{\mathcal{Z}}
\safemath{\emptySet}{\varnothing}

\safemath{\colA}{\mathscr{A}}
\safemath{\colB}{\mathscr{B}}
\safemath{\colC}{\mathscr{C}}
\safemath{\colD}{\mathscr{D}}
\safemath{\colE}{\mathscr{E}}
\safemath{\colF}{\mathscr{F}}
\safemath{\colG}{\mathscr{G}}
\safemath{\colH}{\mathscr{H}}
\safemath{\colI}{\mathscr{I}}
\safemath{\colJ}{\mathscr{J}}
\safemath{\colK}{\mathscr{K}}
\safemath{\colL}{\mathscr{L}}
\safemath{\colM}{\mathscr{M}}
\safemath{\colN}{\mathscr{N}}
\safemath{\colO}{\mathscr{O}}
\safemath{\colP}{\mathscr{P}}
\safemath{\colQ}{\mathscr{Q}}
\safemath{\colR}{\mathscr{R}}
\safemath{\colS}{\mathscr{S}}
\safemath{\colT}{\mathscr{T}}
\safemath{\colU}{\mathscr{U}}
\safemath{\colV}{\mathscr{V}}
\safemath{\colW}{\mathscr{W}}
\safemath{\colX}{\mathscr{X}}
\safemath{\colY}{\mathscr{Y}}
\safemath{\colZ}{\mathscr{Z}}

\safemath{\opA}{\mathbb{A}}
\safemath{\opB}{\mathbb{B}}
\safemath{\opC}{\mathbb{C}}
\safemath{\opD}{\mathbb{D}}
\safemath{\opE}{\mathbb{E}}
\safemath{\opF}{\mathbb{F}}
\safemath{\opG}{\mathbb{G}}
\safemath{\opH}{\mathbb{H}}
\safemath{\opI}{\mathbb{I}}
\safemath{\opJ}{\mathbb{J}}
\safemath{\opK}{\mathbb{K}}
\safemath{\opL}{\mathbb{L}}
\safemath{\opM}{\mathbb{M}}
\safemath{\opN}{\mathbb{N}}
\safemath{\opO}{\mathbb{O}}
\safemath{\opP}{\mathbb{P}}
\safemath{\opQ}{\mathbb{Q}}
\safemath{\opR}{\mathbb{R}}
\safemath{\opS}{\mathbb{S}}
\safemath{\opT}{\mathbb{T}}
\safemath{\opU}{\mathbb{U}}
\safemath{\opV}{\mathbb{V}}
\safemath{\opW}{\mathbb{W}}
\safemath{\opX}{\mathbb{X}}
\safemath{\opY}{\mathbb{Y}}
\safemath{\opZ}{\mathbb{Z}}
\safemath{\opZero}{\mathbb{O}}
\safemath{\identityop}{\opI}

\safemath{\veca}{\bma}
\safemath{\vecb}{\bmb}
\safemath{\vecc}{\bmc}
\safemath{\vecd}{\bmd}
\safemath{\vece}{\bme}
\safemath{\vecf}{\bmf}
\safemath{\vecg}{\bmg}
\safemath{\vech}{\bmh}
\safemath{\veci}{\bmi}
\safemath{\vecj}{\bmj}
\safemath{\veck}{\bmk}
\safemath{\vecl}{\bml}
\safemath{\vecm}{\bmm}
\safemath{\vecn}{\bmn}
\safemath{\veco}{\bmo}
\safemath{\vecp}{\bmp}
\safemath{\vecq}{\bmq}
\safemath{\vecr}{\bmr}
\safemath{\vecs}{\bms}
\safemath{\vect}{\bmt}
\safemath{\vecu}{\bmu}
\safemath{\vecv}{\bmv}
\safemath{\vecw}{\bmw}
\safemath{\vecx}{\bmx}
\safemath{\vecy}{\bmy}
\safemath{\vecz}{\bmz}

\safemath{\veczero}{\bmzero}
\safemath{\vecone}{\bmone}
\safemath{\vecxi}{\bmxi}
\safemath{\veclambda}{\bmlambda}
\safemath{\vecmu}{\bmmu}
\safemath{\vectheta}{\bmtheta}
\safemath{\vecphi}{\bmphi}
\safemath{\vecdelta}{\bmdelta}

\safemath{\matA}{\bA}
\safemath{\matB}{\bB}
\safemath{\matC}{\bC}
\safemath{\matD}{\bD}
\safemath{\matE}{\bE}
\safemath{\matF}{\bF}
\safemath{\matG}{\bG}
\safemath{\matH}{\bH}
\safemath{\matI}{\bI}
\safemath{\matJ}{\bJ}
\safemath{\matK}{\bK}
\safemath{\matL}{\bL}
\safemath{\matM}{\bM}
\safemath{\matN}{\bN}
\safemath{\matO}{\bO}
\safemath{\matP}{\bP}
\safemath{\matQ}{\bQ}
\safemath{\matR}{\bR}
\safemath{\matS}{\bS}
\safemath{\matT}{\bT}
\safemath{\matU}{\bU}
\safemath{\matV}{\bV}
\safemath{\matW}{\bW}
\safemath{\matX}{\bX}
\safemath{\matY}{\bY}
\safemath{\matZ}{\bZ}
\safemath{\matzero}{\bmzero}

\safemath{\matDelta}{\bDelta}
\safemath{\matLambda}{\bLambda}
\safemath{\matPhi}{\bPhi}
\safemath{\matSigma}{\bSigma}
\safemath{\matOmega}{\bOmega}
\safemath{\matTheta}{\bTheta}

\safemath{\matidentity}{\matI}
\safemath{\matone}{\matO}

\safemath{\rnda}{A}
\safemath{\rndb}{B}
\safemath{\rndc}{C}
\safemath{\rndd}{D}
\safemath{\rnde}{E}
\safemath{\rndf}{F}
\safemath{\rndg}{G}
\safemath{\rndh}{H}
\safemath{\rndi}{I}
\safemath{\rndj}{J}
\safemath{\rndk}{K}
\safemath{\rndl}{L}
\safemath{\rndm}{M}
\safemath{\rndn}{N}
\safemath{\rndo}{O}
\safemath{\rndp}{P}
\safemath{\rndq}{Q}
\safemath{\rndr}{R}
\safemath{\rnds}{S}
\safemath{\rndt}{T}
\safemath{\rndu}{U}
\safemath{\rndv}{V}
\safemath{\rndw}{W}
\safemath{\rndx}{X}
\safemath{\rndy}{Y}
\safemath{\rndz}{Z}

\safemath{\rveca}{\bimA}
\safemath{\rvecb}{\bimB}
\safemath{\rvecc}{\bimC}
\safemath{\rvecd}{\bimD}
\safemath{\rvece}{\bimE}
\safemath{\rvecf}{\bimF}
\safemath{\rvecg}{\bimG}
\safemath{\rvech}{\bimH}
\safemath{\rveci}{\bimI}
\safemath{\rvecj}{\bimJ}
\safemath{\rveck}{\bimK}
\safemath{\rvecl}{\bimL}
\safemath{\rvecm}{\bimM}
\safemath{\rvecn}{\bimN}
\safemath{\rveco}{\bomO}
\safemath{\rvecp}{\bimP}
\safemath{\rvecq}{\bimQ}
\safemath{\rvecr}{\bimR}
\safemath{\rvecs}{\bimS}
\safemath{\rvect}{\bimT}
\safemath{\rvecu}{\bimU}
\safemath{\rvecv}{\bimV}
\safemath{\rvecw}{\bimW}
\safemath{\rvecx}{\bimX}
\safemath{\rvecy}{\bimY}
\safemath{\rvecz}{\bimZ}

\safemath{\rvecxi}{\bmxi}
\safemath{\rveclambda}{\bmlambda}
\safemath{\rvecmu}{\bmmu}
\safemath{\rvectheta}{\bmtheta}
\safemath{\rvecphi}{\bmphi}

\safemath{\rmatA}{\bimA}
\safemath{\rmatB}{\bimB}
\safemath{\rmatC}{\bimC}
\safemath{\rmatD}{\bimD}
\safemath{\rmatE}{\bimE}
\safemath{\rmatF}{\bimF}
\safemath{\rmatG}{\bimG}
\safemath{\rmatH}{\bimH}
\safemath{\rmatI}{\bimI}
\safemath{\rmatJ}{\bimJ}
\safemath{\rmatK}{\bimK}
\safemath{\rmatL}{\bimL}
\safemath{\rmatM}{\bimM}
\safemath{\rmatN}{\bimN}
\safemath{\rmatO}{\bimO}
\safemath{\rmatP}{\bimP}
\safemath{\rmatQ}{\bimQ}
\safemath{\rmatR}{\bimR}
\safemath{\rmatS}{\bimS}
\safemath{\rmatT}{\bimT}
\safemath{\rmatU}{\bimU}
\safemath{\rmatV}{\bimV}
\safemath{\rmatW}{\bimW}
\safemath{\rmatX}{\bimX}
\safemath{\rmatY}{\bimY}
\safemath{\rmatZ}{\bimZ}

\safemath{\rmatDelta}{\bimDelta}
\safemath{\rmatLambda}{\bimLambda}
\safemath{\rmatPhi}{\bimPhi}
\safemath{\rmatSigma}{\bimSigma}
\safemath{\rmatOmega}{\bimOmega}
\safemath{\rmatTheta}{\bimTheta}

\usepackage{amssymb}
\usepackage{amsfonts}
\usepackage{mathrsfs}
\usepackage{xspace}
\usepackage{bm}
\usepackage{fancyref}
\usepackage{textcomp}

\usepackage{multirow}
\usepackage{stmaryrd}

\newenvironment{textbmatrix}{	\setlength{\arraycolsep}{2.5pt}%
								\big[\begin{matrix}}{\end{matrix}\big]%
								\raisebox{0.08ex}{\vphantom{M}}}

\def\be{\begin{equation}}
\def\ee{\end{equation}}
\def\een{\nonumber \end{equation}}
\def\mat{\begin{bmatrix}}
\def\emat{\end{bmatrix}}
\def\btm{\begin{textbmatrix}}
\def\etm{\end{textbmatrix}}

\def\ba#1\ea{\begin{align}#1\end{align}}
\def\bas#1\eas{\begin{align*}#1\end{align*}}
\def\bs#1\es{\begin{split}#1\end{split}} 
\def\bg#1\eg{\begin{gather}#1\end{gather}}
\def\bml#1\eml{\begin{multline}#1\end{multline}}
\def\bi#1\ei{\begin{itemize}#1\end{itemize}}

\newcommand{\lefto}{\mathopen{}\left}

\DeclareMathOperator{\tr}{tr}				
\DeclareMathOperator{\sinc}{sinc}			
\DeclareMathOperator{\kron}{\otimes}			
\DeclareMathOperator{\Exop}{\opE}			

\newcommand{\Ex}[2]{\ensuremath{\Exop_{#1}\lefto[#2\right]}} 	
\newcommand{\abs}[1]{\lefto\lvert#1\right\rvert}		

\newcommand{\union}{\cup}					

\newcommand{\opnorm}[1]{\lVert#1\rVert}		

\safemath{\dirac}{\delta}					
\safemath{\krond}{\dirac}					

\safemath{\upto}{\uparrow}
\safemath{\downto}{\downarrow}
\safemath{\iu}{j}							
\safemath{\ev}{\lambda}						
\safemath{\hilseqspace}{l^{2}}				
\newcommand{\banachfunspace}[1]{\setL^{#1}}	
\safemath{\hilfunspace}{\banachfunspace{2}}	

\safemath{\SNR}{\textsf{SNR}} 				
\safemath{\PAR}{\textsf{PAR}} 				
\safemath{\No}{N_0}							
\safemath{\Es}{E_s}							
\safemath{\Eb}{E_b}							
\safemath{\EbNo}{\frac{\Eb}{\No}}
\safemath{\EsNo}{\frac{\Es}{\No}}

\DeclareMathOperator{\CHop}{\ensuremath{\opH}} 
\safemath{\tvir}{\rndh_{\CHop}}				
\safemath{\tvtf}{\rndl_{\CHop}}				
\safemath{\spf}{\rnds_{\CHop}}				
\safemath{\bff}{H_{\CHop}}					

\safemath{\ircf}{r_{h}}						
\safemath{\tftvcf}{r_{s}}					
\safemath{\tfcf}{r_{l}}						
\safemath{\bfcf}{r_{H}}						

\safemath{\tcorr}{c_h}						
\safemath{\scf}{c_{s}}						
\safemath{\tfcorr}{c_{l}}					
\safemath{\fcorr}{c_{H}}						

\safemath{\mi}{I}							
\safemath{\capacity}{C}						

\safemath{\normal}{\mathcal{N}}			
\safemath{\jpg}{\mathcal{CN}}			
\safemath{\mchain}{\leftrightarrow}		

\safemath{\dB}{\,\mathrm{dB}}
\safemath{\dBm}{\,\mathrm{dBm}}
\safemath{\Hz}{\,\mathrm{Hz}}
\safemath{\kHz}{\,\mathrm{kHz}}
\safemath{\MHz}{\,\mathrm{MHz}}
\safemath{\GHz}{\,\mathrm{GHz}}
\safemath{\s}{\,\mathrm{s}}
\safemath{\ms}{\,\mathrm{ms}}
\safemath{\mus}{\,\mathrm{\text{\textmu}s}}
\safemath{\ns}{\,\mathrm{ns}}
\safemath{\ps}{\,\mathrm{ps}}
\safemath{\meter}{\,\mathrm{m}}
\safemath{\mm}{\,\mathrm{mm}}
\safemath{\cm}{\,\mathrm{cm}}
\safemath{\m}{\,\mathrm{m}}
\safemath{\W}{\,\mathrm{W}}
\safemath{\mW}{\, \mathrm{mW}}
\safemath{\J}{\,\mathrm{J}}
\safemath{\K}{\,\mathrm{K}}
\safemath{\bit}{\,\mathrm{bit}}
\safemath{\nat}{\,\mathrm{nat}}

\safemath{\define}{\triangleq}			

\safemath{\equivalent}{\sim}
\safemath{\distas}{\sim}					
\safemath{\sdiff}{\Delta}				

\safemath{\reals}{\mathbb{R}}
\safemath{\positivereals}{\reals_{+}}
\safemath{\integers}{\mathbb{Z}}
\safemath{\posint}{\integers_{+}}
\safemath{\naturals}{\mathbb{N}}
\safemath{\posnaturals}{\naturals_{+}}
\safemath{\complexset}{\mathbb{C}}
\safemath{\rationals}{\mathbb{Q}}

\newcommand*{\fancyrefapplabelprefix}{app}		
\newcommand*{\fancyrefthmlabelprefix}{thm}		
\newcommand*{\fancyreflemlabelprefix}{lem}		
\newcommand*{\fancyrefcorlabelprefix}{cor}		
\newcommand*{\fancyrefdeflabelprefix}{def}		
\newcommand*{\fancyrefproplabelprefix}{prop}	
\newcommand*{\fancyrefobslabelprefix}{obs}		
\newcommand*{\fancyrefalglabelprefix}{alg}		
\newcommand*{\fancyrefasmlabelprefix}{asm}	    

\frefformat{vario}{\fancyrefseclabelprefix}{Sec.~#1}
\frefformat{vario}{\fancyrefthmlabelprefix}{Theorem~#1}
\frefformat{vario}{\fancyreflemlabelprefix}{Lemma~#1}
\frefformat{vario}{\fancyrefcorlabelprefix}{Corollary~#1}
\frefformat{vario}{\fancyrefdeflabelprefix}{Definition~#1}
\frefformat{vario}{\fancyrefobslabelprefix}{Observation~#1}
\frefformat{vario}{\fancyrefasmlabelprefix}{Assumption~#1}
\frefformat{vario}{\fancyreffiglabelprefix}{Fig.~#1}
\frefformat{vario}{\fancyrefapplabelprefix}{Appendix~#1} 
\frefformat{vario}{\fancyrefproplabelprefix}{Proposition~#1}
\frefformat{vario}{\fancyrefalglabelprefix}{Algorithm~#1}
\frefformat{vario}{\fancyrefeqlabelprefix}{(#1)}

\safemath{\dictab}{[\,\dicta\,\,\dictb\,]}

\safemath{\ysig}{\bmy}
\safemath{\ysighat}{\hat{\ysig}}
\safemath{\ysigdim}{M}
\safemath{\xsig}{\bmx}
\safemath{\xsigdim}{N}
\safemath{\nx}{n_x}
\safemath{\zsig}{\bmz}
\safemath{\zsigdim}{\ysigdim}
\safemath{\rsig}{\bmr}
\safemath{\Adict}{\bA}
\safemath{\Adicttilde}{\widetilde{\Adict}}
\safemath{\Adictdim}{\outputdim\times\xsigdim}
\safemath{\avec}{\bma}
\safemath{\avectilde}{\tilde{\avec}}
\safemath{\Bdict}{\bB}
\safemath{\Bdicttilde}{\widetilde{\Bdict}}
\safemath{\Cdict}{\bC}
\safemath{\cvec}{\bmc}
\safemath{\Ddict}{\bD}
\safemath{\Ddictdim}{\ysigdim\times\xsigdim}
\safemath{\dvec}{\bmd}
\safemath{\Ddicttilde}{\widetilde{\bD}}
\safemath{\Bonb}{\bB}
\safemath{\bvec}{\bmb}
\safemath{\Bonbdim}{\ysigdim\times\ysigdim}
\safemath{\noise}{\bmn}
\safemath{\noisedim}{\ysigim}
\safemath{\err}{\bme}
\safemath{\errdim}{\ysigdim}
\safemath{\errset}{\setE}
\safemath{\nerr}{n_e}
\safemath{\delop}{\bP_\errset}
\safemath{\delopc}{\bP_{{\errset}^c}}

\safemath{\cplxi}{\imath}
\safemath{\cplxj}{\jmath}

\safemath{\dict}{\matD}
\safemath{\inputdim}{N}		
\safemath{\outputdim}{M}		
\safemath{\sparsity}{S}	
\safemath{\inputdimA}{{N_a}}	
\safemath{\inputdimB}{{N_b}}	
\safemath{\elemA}{{n_a}}	
\safemath{\elemB}{{n_b}}	
\safemath{\resA}{\matR_a}	
\safemath{\resB}{\matR_b}	
\safemath{\subD}{\matS} 
\safemath{\subA}{\matS_a} 
\safemath{\subB}{\matS_b} 
\safemath{\dicta}{\matA} 	
\safemath{\dictb}{\matB} 	
\safemath{\hollowS}{H}
\safemath{\hollowA}{H_a}
\safemath{\hollowB}{H_b}
\safemath{\cross}{Z}
\safemath{\coh}{\mu_d}			
\safemath{\coha}{\mu_a}			
\safemath{\cohb}{\mu_b}			
\safemath{\mubs}{\nu}	
\safemath{\cohm}{\mu_m} 
\safemath{\dictset}{\setD}	
\safemath{\dictsetp}{\dictset(\coh,\coha,\cohb)}	
\safemath{\dictsetgen}{\dictset_\text{gen}}
\safemath{\dictsetgenp}{\dictsetgen(\coh)}
\safemath{\dictsetonb}{\dictset_\text{onb}}
\safemath{\dictsetonbp}{\dictsetonb(\coh)}

\safemath{\leftside}{U}
\safemath{\rightsideA}{R_a}
\safemath{\rightsideB}{R_b}

\safemath{\indexS}{\setI_S} 

\safemath{\na}{n_a}			
\safemath{\nb}{n_b}			
\safemath{\coeffa}{p_i}	
\safemath{\coeffb}{q_j}	
\safemath{\seta}{\setP}		
\safemath{\setb}{\setQ}     
\safemath{\setw}{\setW}	
\safemath{\setz}{\setZ}	
\safemath{\cola}{\veca}		
\safemath{\colb}{\vecb}		
\safemath{\cold}{\vecd}		
\safemath{\inputvec}{\vecx} 	
\safemath{\error}{\vece}	
\safemath{\noiseout}{\vecz} 	
\safemath{\inputvecel}{x}
\safemath{\inputveca}{\vecx_a}
\safemath{\inputvecb}{\vecx_b}
\safemath{\outputvec}{\vecy}	
\safemath{\lambdamin}{\lambda_{\mathrm{min}}}

\safemath{\elltwo}{\ell_2}
\safemath{\ellone}{\ell_1}
\safemath{\ellzero}{\ell_0}
\safemath{\ellinf}{\ell_\infty}
\safemath{\ellinftilde}{\ell_{\widetilde\infty}}
\safemath{\licard}{Z(\coh,\coha,\cohb)}
\safemath{\xsol}{\hat{x}}
\safemath{\xbord}{x_b}		
\safemath{\xstat}{x_s}		
\safemath{\xstatLone}{\tilde{x}_s}
\safemath{\order}{\mathcal{O}} 
\safemath{\scales}{\Theta} 
\safemath{\ones}{\mathbf{1}} 
\safemath{\zeroes}{\mathbf{0}} 
\safemath{\thlone}{\kappa(\coh,\cohb)} 
\safemath{\constoneA}{\delta} 
\safemath{\constoneB}{\epsilon} 
\safemath{\nlarge}{L}				   
\safemath{\sumlarge}{S_\nlarge}
\safemath{\maxlarger}{P_\nlarge}	   
\safemath{\Pzero}{\textrm{P0}}	
\safemath{\Pone}{\textrm{P1}}
\safemath{\vecfir}{\vecw}			 
\safemath{\vecsec}{\vecz}
\safemath{\elvecfir}{w}              
\safemath{\elvecsec}{z}				 
\safemath{\nlargefir}{n}
\safemath{\normout}{\gamma}
\safemath{\auxfun}{h}
\safemath{\supp}{\textrm{supp}}

\safemath{\indexa}{\ell}
\safemath{\indexb}{r}
\safemath{\indexc}{i}
\safemath{\indexd}{j}

\safemath{\project}{P}

\newcommand{\quantize}{\mathcal{Q}}

\DeclareMathOperator{\diag}{diag}

\makeatletter
\def\@IEEEinterspaceratioM{0.265}
\def\@IEEEinterspaceMINratioM{0.1651}
\def\@IEEEinterspaceMAXratioM{0.38}

\def\@IEEEinterspaceratioB{0.31}
\def\@IEEEinterspaceMINratioB{0.19}
\def\@IEEEinterspaceMAXratioB{0.38}
\@IEEEtunefonts
\makeatother
\begin{document}

%

\title{On Out-of-Band Emissions of Quantized Precoding in Massive MU-MIMO-OFDM} 
\author{
\IEEEauthorblockN{Sven Jacobsson$^\text{1,2,3}$, Giuseppe Durisi$^\text{2}$, Mikael Coldrey$^\text{1}$, and Christoph Studer$^\text{3}$} \\ \vspace{-0.3cm}          
\IEEEauthorblockA{$^\text{1}$Ericsson Research, Gothenburg, Sweden}
\IEEEauthorblockA{$^\text{2}$Chalmers University of Technology, Gothenburg, Sweden}
\IEEEauthorblockA{$^\text{3}$Cornell University, Ithaca, NY, USA} 
\vspace{-0.3cm}
\thanks{The work of SJ and GD was supported in part by the Swedish Foundation for Strategic Research under grant ID14-0022, and by the Swedish Governmental Agency for Innovation Systems (VINNOVA) within the competence center ChaseOn. SJ's research visit to Cornell was sponsored in part by Cornell's College of Engineering. The work of CS was supported by Xilinx, Inc.~and by the US National Science Foundation~(NSF) under grants ECCS-1408006, CCF-1535897, CAREER CCF-1652065, and CNS-1717559.}
}

\maketitle


\begin{abstract}
We analyze out-of-band (OOB) emissions in the massive multi-user (MU) multiple-input multiple-output~(MIMO) downlink. 
We focus on systems in which the base station~(BS) is equipped with low-resolution digital-to-analog converters~(DACs) and orthogonal frequency-division multiplexing~(OFDM) is used to communicate to the user equipments (UEs) over frequency-selective channels.
We demonstrate that analog filtering in combination with simple frequency-domain digital predistortion~(DPD) at the BS enables a significant reduction of OOB emissions, but degrades the signal-to-interference-noise-and-distortion ratio~(SINDR) at the UEs and increases the peak-to-average power ratio (PAR) at the BS.  
We use Bussgang's theorem to characterize the tradeoffs between OOB emissions, SINDR, and PAR, and to study the impact of analog filters and DPD on the error-rate performance of the massive MU-MIMO-OFDM downlink. 
Our results show that by carefully tuning the parameters of the analog filters, one can achieve a significant reduction in OOB emissions with only a moderate degradation of error-rate performance and PAR. 
%
%
\end{abstract}

\section{Introduction}


Massive multi-user (MU) multiple-input multiple-output (MIMO) in 
combination with orthogonal frequency-division multiplexing (OFDM) is widely believed to be among the key technologies in fifth-generation (5G) cellular systems~\cite{boccardi14a}.
However, a  base-station (BS) with digital beamforming capabilities and hundreds of  radio-frequency (RF) chains must necessarily rely on low-cost (and, hence, low-quality) RF components. 
In this paper, we focus on BS designs that use low-resolution digital-to-analog converters (DACs) in the downlink. Such architectures  promise significant reductions in costs and circuit power consumption, and provide means to lower the interconnect bandwidth between the baseband processor and the radio unit. 

\subsection{Previous Work}
For frequency-flat channels and single-carrier transmission, it has been shown in~\cite{mezghani09c,saxena16b,li17a,jacobsson17d} that linear precoding followed by coarse quantization enables low uncoded bit-error rates~(BERs) and high achievable rates in massive MU-MIMO.
For the extreme case of 1-bit DACs, it was shown in~\cite{jacobsson17d, jedda16a, castaneda17a, swindlehurst17a} that more sophisticated nonlinear precoders can further improve the system performance.
For frequency-selective channels, the performance achievable by using OFDM in combination with 1-bit DACs was recently studied in~\cite{guerreiro16a, jacobsson17c, jacobsson17f}. 
For linear precoding, the analysis in~\cite{jacobsson17c} was extended in~\cite{jacobsson17e} to multi-bit~DACs.

All the above results ignore that the use of low-resolution DACs cause unwanted out-of-band (OOB) emissions, which need to be mitigated for the resulting transmit waveform to satisfy the spectral requirements imposed by regulatory bodies.
To our knowledge, the only work to consider analog filtering combined with low-resolution DACs is~\cite{jedda15b}, which deals with the design of pulse-shaping filters and investigates bandwidth efficiency for oversampled 1-bit-DACs and single-carrier transmission.
OOB emissions caused by nonlinear power amplifiers (PAs) at the BS are analyzed in~\cite{mollen16b, mollen16e}.
There, OOB emissions are measured in terms of the adjacent channel leakage ratio (ACLR), i.e., the ratio between the power leaked to adjacent frequency bands (due to hardware impairments) and the in-band power. 
It has been noted in~\cite{mollen16b} that OOB emissions may not be problematic in massive MU-MIMO because the total transmit power is significantly lower than in traditional small-scale MIMO systems. 

\subsection{Contributions}
In this paper, we investigate the OOB emissions in the massive MU-MIMO-OFDM downlink caused by low-resolution DACs. We extend the Bussgang-based analysis from~\cite{jacobsson17c, jacobsson17e} to a more accurate transmitter model that includes analog filtering after the quantizer in the DAC. 
We show how analog filtering combined with digital predistortion (DPD) can be used to mitigate, to some extent, the OOB emissions caused by low-resolution DACs. 
We also investigate the spectral and spatial characteristics of the distortion caused by low-resolution DACs. 
Finally, we study the tradeoffs between ACLR at the BS, the peak-to-average power ratio~(PAR) at the~BS, and signal-to-interference-noise-and-distortion ratio (SINDR) at the UEs.

\subsection{Notation}

%
%
%
The $M \times N$ all-zeros matrix and the $M \times M$ identity matrix are denoted by $\veczero_{M \times N}$ and $\matI_M$, respectively.
%
%
The real and the imaginary parts of a complex-valued vector $\veca$ are $\Re\{\veca\}$ and $\Im\{\veca\}$, respectively. 
The $\ell_{\widetilde\infty}$-norm of~$\veca = [a_1, \dots, a_M]^T$ is $\opnorm{\veca}_{\widetilde\infty} = \max\big\{ \|\Re\{\bma\}\|_\infty, \|\Im\{\bma\}\|_\infty \big\}$, where $\opnorm{\veca}_\infty = \max_{m=1,\ldots,M}  \abs{a_m}$.
We use $\opnorm{\veca}_2$ to denote the $\ell_2$-norm of $\veca$. 
The matrix $\diag(\veca)$ is diagonal with the vector~$\veca$ on its main diagonal. 
If $\matA$ is an $M \times N$ matrix, then $\text{vec}(\matA)$ is a $MN$-dimensional vector obtained by column-wise stacking of the columns of $\matA$. 
%
%
%
The complex-valued circularly symmetric Gaussian distribution with covariance matrix $\matK \in \opC^{M \times M}$ is~$\jpg(\matzero_{M \times 1}, \matK)$.
The uniform distribution on the interval $(a,b)$ is $\mathcal{U}(a,b)$.
%
%
The indicator function is defined as $\mathds{1}_\setA(a) = 1$ for $a \in \setA$ and $\mathds{1}_\setA(a) = 0$ for $a \notin \setA$. Finally, we define $\sinc(x) = \sin(\pi x) / (\pi x)$.


\section{A Simple DAC Model}

\setlength{\textfloatsep}{10pt}
\begin{figure}
\begin{center}
	\includegraphics[width = \columnwidth]{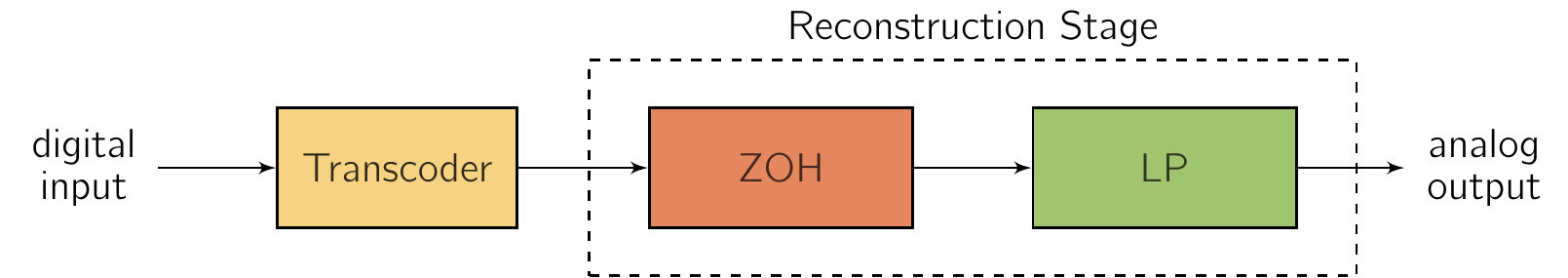}	
	\caption{Block diagram of the basic components of a DAC~\cite[Fig.~1.1]{maloberti07a}.}
	\label{fig:dac}
\end{center}
\end{figure}

We consider a BS equipped with $B$ antennas where two identical DACs at each antenna generate the in-phase and quadrature components of the transmitted signal.
Each DAC is modeled as illustrated in~\fref{fig:dac} and performs two operations: the \emph{transcoder} generates an analog sequence whose amplitude is the analog representation of the digital input, and the \emph{reconstruction stage} maps the transcoder output to a continuous-time waveform. 
Practical reconstruction stages commonly cascade a zero-order hold~(ZOH) filter and a low-pass (LP) filter~\cite[Sec.~1.7]{maloberti07a}.

\subsection{Quantization}
Let $z_{b,n}$ denote the $n$th sample ($n = 0, \dots, N-1$) of the digital input at the $b$th antenna ($b = 1, \dots, B$).
We assume that the digital input has infinite precision, i.e., that $z_{b,n} \in \opC$, whereas the transcoder in the DACs supports only~$2^Q$ voltage levels, with $Q$ being the number of DAC bits. Therefore, we model the transcoder as a \emph{quantizer}, i.e., a nonlinear function $\quantize(\cdot)$ that, at each sampling instant $n$, maps a sample in~$\opC$ to a quantized sample belonging to the finite-cardinality set  $\{ q_0, \dots,q_{2^Q-1}\}\times\{ q_0, \dots,q_{2^Q-1}\}$ as follows: $\quantize(z_{b,n}) = \sum_{i=0}^{2^Q-1} q_i \mathds{1}_{[\tau_i, \tau_{i+1})}\lefto(\Re\{z_{b,n}\}\right) + j\sum_{i=0}^{2^Q-1} q_i \mathds{1}_{[\tau_i, \tau_{i+1})}\lefto(\Im\{z_{b,n}\}\right)$. 
We consider symmetric uniform quantizers for which the quantization \emph{labels} are $q_i = \alpha\Delta(i - (2^Q-1)/2)$ for $i = 0,\dots,2^Q-1$, and the quantization \emph{thresholds} are $\tau_i = \Delta(i - 2^Q/2)$ for $i = 1,\dots,2^Q-1$ with $\tau_0 = -\infty$ and $\tau_{2^Q} = \infty$. Here, $\Delta$ is the \emph{step size} of the DACs and $\alpha$ is a constant ensuring that the  power constraint  $\Ex{}{\lvert\quantize(z_{b,n})\rvert^2} = (B\cdot\textit{OSR})^{-1}$ is satisfied, where $\textit{OSR}$ is the oversampling ratio (OSR) of the DACs~(see~\fref{sec:ofdm}).

\subsection{Reconstruction}

Let $f_\text{s}$ and $T_\text{s} = 1/f_\text{s}$ denote the sampling rate and the sampling period of the DACs, respectively.
We assume that the bandwidth of the desired (in-band) signal is contained within the interval $(-f_\text{BW}/2, f_\text{BW}/2)$, where $f_\text{BW} \le f_\text{s}$.
The spectrum of the input to the DACs is periodic with replicas of the in-band signal centered around integer multiples of~$f_\text{s}$.
%
%
Hence, an ideal reconstruction stage would be a LP filter with frequency response $\hat{r}_\text{LPF}(f) = \mathds{1}_{(-f_\text{cut}/2,f_\text{cut}/2)}(f)$, where $f_\text{BW}/2 \le f_\text{cut} \le f_\text{s}/2$. 
Indeed, the setup studied in~\cite{mezghani09c, saxena16b, li17a, jacobsson17d, jedda16a, castaneda17a, guerreiro16a, jacobsson17c, swindlehurst17a, jacobsson17e, jacobsson17f} requires implicitly that such an LP filter (with $f_\text{cut} = f_\text{s}/2$) is used.
Unfortunately, ideal LP filters cannot be realized in practice since the corresponding impulse response is noncausal and has infinite support. Therefore, we consider a ZOH filter followed by a nonideal (but practical) LP filter in the reconstruction stage. 

%

%
%
The ZOH filter holds each sample value for one sampling period. The impulse response of the ZOH filter is $r_\text{ZOH}(t) = T_\text{s}^{-1} \mathds{1}_{[0,T_\text{s})}(t)$ and the corresponding frequency response is
\begin{align} \label{eq:zoh_freq}
	\hat{r}_\text{ZOH}(f) &= \int_{-\infty}^\infty r_\text{ZOH}(t) e^{-2\pi f t} \text{d}t = e^{-j \pi f T_\text{s}} \sinc(f T_\text{s}). 
\end{align}
We note that the ZOH filter is a LP filter---a desirable feature because it attenuates the OOB emissions, which consist of quantization noise and replicas of the in-band signal. 
We also assume that a dedicated LP filter is cascaded with the ZOH filter to further reduce OOB emissions. 
We consider a Butterworth filter of order $\eta \in \{ 1, 2\}$, for which the frequency response is
\begin{align}
\hat{r}_\text{LP}(f) &= 
\begin{cases}
(1 + jf/f_\text{cut})^{-1}, & \eta = 1,	 \\
\lefto(1 + j\sqrt{2}f/f_\text{cut} - (f/f_\text{cut})^2\right)^{-1}, & \eta = 2.	
\end{cases} \label{eq:lpf_freq}
\end{align}
After the reconstruction stage (i.e., after the ZOH filter and the LP filter), the continuos-time continuos-amplitude DAC output at the $b$th antenna ($b = 1,\dots, B$) can be written as
\begin{IEEEeqnarray}{rCl}
	x_{b}(t) &=& \frac{1}{T_\text{s}}\!\sum_{n = 0}^{N-1}\int_{-\infty}^\infty \! r_\text{LP}(t-u) \quantize(z_{b,n}) \mathds{1}_{T_\text{s}[n, n+1)}(u) \, \text{d}u \IEEEeqnarraynumspace
\end{IEEEeqnarray}
where $r_\text{LP}(t) = \int_{-\infty}^\infty \hat{r}_\text{LP}(f)e^{j2\pi f t} \text{d}f$.

\section{Massive MU-MIMO-OFDM Downlink} \label{sec:ofdm}

We consider a single-cell massive MU-MIMO-OFDM downlink scenario in which a $B$-antenna BS serves $U$ single-antenna UEs in the same time-frequency resource.
We assume that OFDM is used to simplify equalization when communicating over frequency-selective channels. 
Specifically, we denote by~$S$ the number of occupied subcarriers per OFDM symbol, and by $N$ the number of time samples per OFDM symbol (which coincides with the total number of available subcarriers).
The set of occupied subcarriers is $\setS = \{ 1,\dots,S/2, N - S/2,\dots, N-1\}$ and the set of unused subcarriers is $\setG = \{ 0, \dots, N-1\} \setminus \setS$.
Let $\vecs_k$ denote the $U$-dimensional symbol vector corresponding to the $k$th subcarrier. 
We shall assume that $\vecs_k \sim \jpg(\veczero_{U \times 1}, \matI_U)$ for $k \in \setS$ and that $\vecs_k \in \veczero_{U \times 1}$ for $k \in \setG$.
We define the OSR as $\textit{OSR} = N/S$.
At each BS antenna, a cyclic prefix (CP) is prepended to  the transmitted signal.
We assume that the length of the CP exceeds the effective length of the LP filter and of the impulse response of the multipath channel.

\subsection{Channel Input-Output Relation}

We assume that the sampling rate of the ADCs at the UEs equals the sampling rate $f_\text{s}$ of DACs at the BS and that the ADCs have infinite resolution (no quantization is used). Also, the anti-aliasing filter in the ADCs is an ideal LP filter with cut-off frequency $f_\text{s}/2$.
Under these assumptions, the continuous-amplitude discrete-time baseband signal $\vecy_n$  at the $U$ UEs~is
\begin{IEEEeqnarray}{rCl}
	\vecy_n &=& \sum_{\ell = 0}^{L-1} \matH_\ell\vecx_{n-\ell} + \vecw_n, \quad n = 0,\dots, N-1.
\end{IEEEeqnarray}
Here, $\vecx_n = [x_{1,n},\dots, x_{B,n}]^T \in \opC^B$ is the DAC output (after the reconstruction stage) sampled at time instant $nT_\text{s}$, i.e., $x_{b,n} = x_b(nT_\text{s})$.
Furthermore, $\vecw_n \sim \jpg(\veczero_{U \times 1}, N_0\matI_U)$ is the AWGN at the UEs, and $\matH_\ell \in \opC^{U \times B}$ is the $\ell$th tap of the frequency-selective MIMO channel matrix ($\ell = 0,1,\dots,L-1$). The realizations of $\{ \matH_\ell \}_{\ell=0}^{L-1}$ are assumed to be known to the BS and to remain constant for the duration of an OFDM symbol (including the duration of the CP). 
The corresponding frequency-domain received vector at the $k$th subcarrier~is
\begin{IEEEeqnarray}{rCl} \label{eq:inout_freq}
	\hat\vecy_k &=& \widehat\matH_k \hat\vecx_k + \hat\vecw_k, \quad k = 0, \dots, N-1.
\end{IEEEeqnarray}
Here, $\widehat\matH_k = \sum_{\ell = 0}^{L-1}\matH_\ell e^{-jk\frac{2\pi}{N}\ell}$. Furthermore, $\hat\vecx_k$, $\hat\vecy_k$, and $\hat\vecw_k$ is the $k$th column of the $B \times N$ matrices $[\vecx_0, \dots, \vecx_{N-1}]\matF$, $[\vecy_0, \dots, \vecy_{N-1}]\matF$, and $[\vecw_0, \dots, \vecw_{N-1}]\matF$, respectively. Here,~$\matF$ is the $N \times N$ unitary discrete Fourier transform (DFT) matrix. 


\subsection{Channel Model}

We consider a simple plane-wave model in which there is one strong line-of-sight (LoS) path and several non-LoS (nLoS) paths from the BS to each UE. 
The BS antennas form a uniform linear array (ULA) with half-wavelength spacing. 
The element on the $u$th row and on the $b$th column of the $U \times B$ channel matrix associated with the $\ell$th tap ($\ell = 0, \dots, L-1$)~is 
\begin{IEEEeqnarray}{rCl}
	\lefto[\matH_{\ell}\right]_{u,b} &=&  \psi_u\gamma_{u,\ell} e^{-j\pi(b-1)\cos(\theta_{u,\ell})}.
\end{IEEEeqnarray}
Let~$\phi_{u}$ denote the angle-of-departure (AoD) from the BS to the $u$th UE. We assume that $\theta_{u,0} = \phi_{u}$ (LoS path) and that $\theta_{u,\ell} = \phi_u + \theta_\ell$ where $\theta_\ell \sim \setU[-180^\circ, 180^\circ]$, for $\ell = 1, \dots, L-1$ (nLoS paths).
The large-scale fading is modeled via $\psi_u^2 = (100/\delta_u)^2$ (free-space path loss), where $\delta_u$ is the distance (in meters) from the BS to the $u$th UE.
For the LoS path, we set $\gamma_{u, 0}^2 = 3/4$; for the nLoS paths, we assume an exponential power delay profile.
Specifically, $\gamma_{u,\ell}^2 \propto  e^{-\ell}$ for $\ell = 1, \dots, L-1$ and  $\sum_{\ell=1}^{L-1} \gamma_{u,\ell}^2 = 1/4$ for $u = 1, \dots, U$.

\subsection{Linear Precoding and Predistortion}

It turns out to be convenient to compactly represent the frequency-domain input-output relation~\eqref{eq:inout_freq} as
\begin{IEEEeqnarray}{rCl} \label{eq:inout_vector}
	\hat\vecy &=& \widehat\matH\hat\vecx	 + \hat\vecw. 
\end{IEEEeqnarray}
Here, $\hat\vecx = \text{vec}\lefto([\hat\vecx_0, \dots, \hat\vecx_{N}]\right)$, $\hat\vecy = \text{vec}\lefto([\hat\vecy_0, \dots, \hat\vecy_{N}]\right)$, $\hat\vecw = \text{vec}\lefto([\hat\vecw_0, \dots, \hat\vecw_{N}]\right)$, and $\widehat\matH \in \opC^{B N \times B N}$ is the block-diagonal matrix having the matrices $\widehat\matH_{0},\dots,\widehat\matH_{N-1}$ on the main~diagonal. 
%
We further write $\hat\vecx$ as
\begin{IEEEeqnarray}{rCl} \label{eq:awesome}
\hat\vecx &=&  \big( \diag(\hat\vecr)\matF \kron \matI_B \big) \quantize\lefto( \vecz \right). \IEEEeqnarraynumspace
\end{IEEEeqnarray}
Here, $\vecz = \text{vec}\lefto( [\vecz_0, \dots, \vecz_{N-1}] \right)$, where $\vecz_n = [z_{1,n}, \dots, z_{B,n}]^T$ is the DAC input at discrete time $n$.
Furthermore, $\hat\vecr = [\hat{r}_0,\dots,\hat{r}_{N-1}]^T$ is the (sampled) frequency response of the analog filters. Specifically, $\hat{r}_k = \hat{r}_\text{LP}(p(k)\Delta{f})\hat{r}_\text{ZOH}(p(k)\Delta{f})$, where $p(k) = (k + N/2) \mod{(N)} - N/2$ and $\Delta{f} = f_\text{s}/N$ is the subcarrier spacing. 
The analog filters in \eqref{eq:zoh_freq} and~\eqref{eq:lpf_freq} do not only reduce OOB emissions, they also attenuate the in-band signal.
To compensate for this attenuation, we assume that the DAC input is \emph{predistorted} in the frequency domain, i.e.,~we try to invert the analog filters in the digital domain. We refer to this approach as DPD.
Specifically, the precoded vector on the $k$th subcarrier is multiplied by $r_k^{-1}=\hat{r}^{-1}_\text{LP}(p(k)\Delta{f})\hat{r}^{-1}_\text{ZOH}(p(k)\Delta{f})$. Hence, after linear precoding and DPD, the vectorized DAC input $\vecz$ can be written as follows:
\begin{IEEEeqnarray}{rCl} \label{eq:even_more_awesome}
	\vecz &=& \big(\matF^H\xi\diag(\hat\vecr)^{-1} \kron \matI_B \big) \widehat\matP\vecs.
\end{IEEEeqnarray}
Here, $\vecs = \text{vec}([\vecs_0, \dots, \vecs_{N-1}])$ and $\widehat\matP \in \opC^{B N \times B N}$ is the block-diagonal matrix with the matrices $\widehat\matP_{0},\dots,\widehat\matP_{N-1}$ on the main~diagonal, where $\widehat\matP_{k} \in \opC^{B \times U}$ is the frequency-domain \emph{precoding matrix} associated with the $k$th subcarrier ($k = 0, \dots, N-1$). 
In what follows, we consider zero-forcing~(ZF) precoding, for which it holds~that
\begin{IEEEeqnarray}{rCl}
	\widehat\matP_k &=& \frac{1}{\sqrt{\frac{1}{S} \sum_{k \in \setS} \tr\bigl( \widehat\matH_{k} \widehat\matH_{k}^H\big)^{-1}}}  \,\widehat\matH_k^H(\widehat\matH_k\widehat\matH_k^H)^{-1}\vecs_k \IEEEeqnarraynumspace
\end{IEEEeqnarray}
for $k \in \setS$.
We use the convention that $\widehat\matP_k = \matzero_{B \times U}$ for $k \in \setG$.
To preserve the transmit power constraint, we rescale the predistorted signal~by
\begin{IEEEeqnarray}{rCl}
\xi &=& 	\sqrt{{\sum_{k \in \setS} \tr\lefto(\widehat\matP_k\widehat\matP_k^H \right)} \Big/  {\sum_{k \in \setS} \tr\lefto(\hat{r}_k^{-1}\widehat\matP_k \big(\hat{r}_k^{-1}\widehat\matP_k\big)^H \right)}}\,. \IEEEeqnarraynumspace
\end{IEEEeqnarray}
Note that $\xi \le 1$ since $\abs{r_k} = \abs{\hat{r}_\text{LP}(p(k)\Delta{f})\hat{r}_\text{ZOH}(p(k)\Delta{f})} \le 1$ for all~$k$.
This confirms that the ZOH and the LP filters attenuate not only the OOB emissions but also the in-band signal, which leads to a power loss even in the infinite-resolution case. 
One can make this loss negligible by operating the DACs at a high OSR (which reduces the attenuation of the in-band signal caused by the ZOH filter) and by increasing $f_\text{cut}$ (which reduces the attenuation of the in-band signal caused by the LP filter). 

\setlength{\textfloatsep}{10pt}
\begin{figure*}[tp]
\centering
\subfloat[PSD of the predistorted DAC input.]{\includegraphics[width=.31\textwidth]{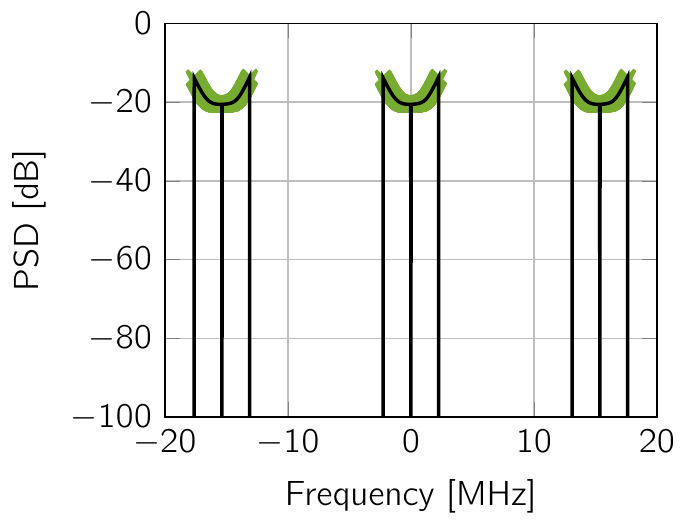}\label{fig:spectrum_before}} \quad
\subfloat[PSD of the 1-bit DAC output.]{\includegraphics[width=.31\textwidth]{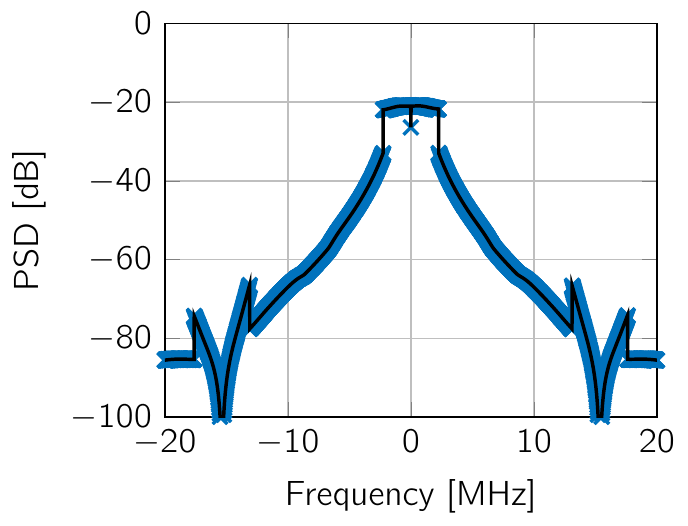}\label{fig:spectrum_1bit}} \quad
\subfloat[PSD of the 3-bit DAC output.]{\includegraphics[width=.31\textwidth]{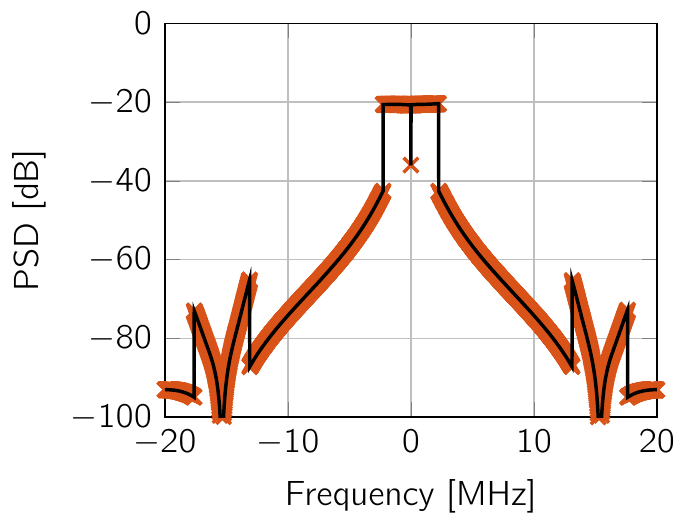}\label{fig:spectrum_3bit}} \\
\caption{PSD of the DAC input and output; $\eta = 2$ and $f_\text{cut} = 1.6875$\,MHz. The markers correspond to simulated values and the solid lines to analytical results.}
\label{fig:spectrum}
\end{figure*}

\subsection{Linearization through Bussgang's Theorem}%
%
%
If the limited resolution of the DACs is accounted for, then the nonlinear nature of $\quantize(\cdot)$ causes intercarrier interference, which renders an exact performance analysis difficult.
To enable an analytic investigation, we follow~\cite{jacobsson17e} and linearize the input-output relation using Bussgang's theorem~\cite{bussgang52a, rowe82a}.
%
%
Specifically, by inserting~\eqref{eq:even_more_awesome} into \eqref{eq:awesome} and by proceeding analogously to~\cite[Sec.~III]{jacobsson17e}, we obtain
\begin{IEEEeqnarray}{rCl} \label{eq:linearized}
\hat\vecx &=&  \lefto(\diag(\hat\vecr)\matF \kron \matI_B \right) \! \big(\matG \lefto( \matF^H\xi\diag(\hat\vecr)^{-1}\! \kron \matI_B \right) \! \widehat\matP\vecs + \vecd \big) \IEEEeqnarraynumspace \label{eq:linear}
\end{IEEEeqnarray}
where the \emph{distortion} $\vecd \in \opC^{BN}$ is uncorrelated with the symbol vector $\vecs$, and where $\matG = \matI_N \otimes \diag(\vecg)$ with
\begin{multline} \label{eq:gainmatrix_uniform}
\text{diag}\lefto( \vecg \right) = \frac{\alpha\Delta}{\sqrt{\pi}} \, \text{diag}\lefto( \matC_{\vecz_n} \right)^{-1/2} \nonumber\\
 \times\sum_{i=1}^{2^Q-1} \! \exp\lefto(-\Delta^2\lefto( i - 2^{Q-1} \right)^2 \text{diag}\lefto( \matC_{\vecz_n} \right)^{-1}\right). 
\end{multline}
%
%
Here, $\matC_{\vecz_n} = ({\xi^2}/{N}) \sum_{k \in \setS} \hat{r}_k^{-1}\widehat\matP_k \big(\hat{r}_k^{-1}\widehat\matP_k\big)^H\!$.  
By inserting~\eqref{eq:linearized} into~\eqref{eq:inout_vector} we find a linear relationship between the transmitted symbol vector $\vecs$ and the frequency-domain received vector~$\hat\vecy$.
Let $\textit{SINDR}_{u,k}\big(\widehat\matH\big)$ denote the SINDR on the $k$th subcarrier for the $u$th UE. 
We find, from~\eqref{eq:inout_vector} and \eqref{eq:linearized}, that
\begin{IEEEeqnarray}{rCl} \label{eq:sindr}
\textit{SINDR}_{u,k}\big(\widehat\matH\big) &=& 	\frac{ \xi^2\lvert\hat\vech_u^T\text{diag}(\vecg)\hat\vecp_u\rvert^2 }{\xi^2\!\!\sum\limits_{v \neq u}\lvert\hat\vech_u^T\text{diag}(\vecg)\hat\vecp_v\rvert^2 \! + \!  D_{u,k}\big(\widehat\matH\big) \! + \!N_0} \IEEEeqnarraynumspace \vspace{-.1cm}
\end{IEEEeqnarray}
where $D_{u,k}(\widehat\matH)$ is the $(u + kU)$th element on the main diagonal of the matrix $\widehat\matH\lefto(\diag(\hat\vecr)\matF\otimes\matI_B \right)\matC_\vecd\lefto(\matF^H\!\diag(\hat\vecr)^H\!\otimes\matI_B \right)\widehat\matH^H$, with $\matC_\vecd = \matC_{\quantize(\vecz)} - \matG\matC_\vecz\matG$ being the covariance of $\vecd$,~and
\begin{IEEEeqnarray}{rCl}
\matC_\vecz &=& \lefto(\matF^H\xi\diag(\hat\vecr)^{-1}\otimes\matI_B \right)\widehat\matP \nonumber\\
&& \quad \times \widehat\matP^H\lefto(\lefto(\xi\diag(\hat\vecr)^{-1}\right)^H\matF\otimes\matI_B\right).		
\end{IEEEeqnarray}
For 1-bit DACs, the covariance $\matC_{\quantize(\vecz)}$ of $\quantize(\vecz)$ can be computed exactly using the so-called \emph{arcsine law}~\cite{van-vleck66a} as~follows:
\begin{IEEEeqnarray}{rCl} 
\matC_{\quantize(\vecz)} &=& \frac{2 S}{\pi B N} \lefto( \arcsin\lefto( \matK_\text{re} \right)  + j\arcsin\lefto( \matK_\text{im} \right) \right). \label{eq:Cxx_arcsine}
\end{IEEEeqnarray}
Here, $\matK_\text{re} =  \text{diag}(\matC_{\vecz})^{-{1}/{2}} \, \Re\{ \matC_{\vecz} \} \, \text{diag}(\matC_{\vecz})^{-{1}/{2}}$ and $\matK_\text{im} =  \text{diag}(\matC_{\vecz})^{-{1}/{2}} \, \Im\{ \matC_{\vecz} \} \, \text{diag}(\matC_{\vecz})^{-{1}/{2}}$
Unfortunately, no closed-form expression for $\matC_{\quantize(\vecz)}$ is available in the multi-bit case, but accurate approximations are provided in~\cite[Sec.~IV]{jacobsson17e}.\footnote{For the parameters considered in this paper, both approximations proposed in~\cite[Sec.~IV]{jacobsson17e} yield similar results.}
%
%

\section{Numerical Results}

We will now present numerical simulation results  and compare them with our analytic results based on Bussgang's theorem linearization. 
Due to space constraints, we shall focus on a selected set of simulation parameters.
We consider (unless stated otherwise) a BS with $B = 64$ antennas serving $U = 4$ UEs. 
We focus on an LTE-inspired scenario: OFDM with $S = 300$ occupied subcarriers; 
the subcarrier spacing is $\Delta{f} = 15$\,kHz and the number of samples per OFDM symbol is $N=1024$; the occupied bandwidth is $f_\text{BW} = S\Delta{f} = 4.5$~MHz, the sampling rate of the DACs is $f_\text{s} = N\Delta{f} = 15.36$~MHz, and  the OSR is $\textit{OSR} = N/S \approx 3.4$.\footnote{In the numerical simulations, the measurement bandwidth is set to $10 f_\text{s} = 153.6$~MHz, i.e., we use $10$ samples to represent the ZOH filter.} 
A CP of length $4.7\,\mu$s ($72$ samples) is prepended to each OFDM symbol.
%
%
We fix $[\phi_1, \phi_2, \phi_3, \phi_4] = [25^\circ, 55^\circ, 75^\circ, 100^\circ]$ and $[\delta_1, \delta_2, \delta_3, \delta_4] = [90, 65, 115, 150]$ meters. 
The number of taps is set to $T=10$ (one \emph{fixed} LoS path and 9 random nLoS~paths per UE). The curves depicted in the figures are obtained by averaging over $100$ random channel realizations (i.e., random realizations of the $9$ nLoS taps).

In the numerical simulations, we  draw the elements of $\vecs_k$ for $k \in \setS$ randomly from a quadrature phase-shift keying (QPSK) constellation, whereas our analytic results assume Gaussian inputs (which is required by Bussgang's theorem).
As we shall see, this assumption results in a negligible performance difference. 
Indeed, each per-antenna DAC input, which is the sum of $US=1200$ independent and identically distributed random variables, is well-approximated by a Gaussian random variable due to the central limit theorem. 
 %

\subsection{Spectral and Spatial Emissions}

In \fref{fig:spectrum}, we plot the power spectral density (PSD) of the DAC input and output (averaged over the BS antennas and the channel realizations) for the case when the LP filter is a second-order Butterworth filter with $f_\text{cut} = 1.6875$~MHz. 
In~\fref{fig:spectrum_before}, we show the spectrum of the predistorted DAC input. Note that $f_\text{cut} < f_\text{BW}/2$, which implies that part of the in-band spectrum is significantly attenuated by the LP filter, which explains the shape of the spectrum in~\fref{fig:spectrum_before} (the analog filters are inverted by the DPD in the digital domain).
We see that replicas of the in-band signal are centered around integer multiples of $f_\text{s} = 15.36$~MHz.

In \fref{fig:spectrum_1bit} and \fref{fig:spectrum_3bit}, we show the spectrum of the DAC output with 1-bit DACs and 3-bit DACs, respectively. 
We first note that the distortion caused by the quantizer is not contained within the transmission bandwidth (i.e., we have indeed OOB emissions), and that the distortion (in-band and OOB) caused by the finite-resolution DACs decreases as the number of~bits increases.
For example, by using 3-bit DACs instead of 1-bit DACs, we see that the PSD on the unoccupied DC carrier is almost $10$~dB lower. 
Note also that the replicas at~$\pm f_\text{s}$ are significantly attenuated by the analog filters. 
The spectrum of the in-band signal after the DACs is (almost) flat (recall that the curves are obtained by averaging over many channel realizations), which shows that the DPD is able to approximately invert the analog filters for $k \in \setS$, despite the coarse quantization. 

Numerical simulations are compared with our analytic results based on Bussgang's theorem linearization.
%
%
Specifically, the PSD on the $b$th BS antenna  at frequency $p(k)\Delta{f}$ is given by $P_{\hat{x}_{b,k}} = [\matC_{\hat\vecx}]_{b + kB, b + kB}$ where $\matC_{\hat\vecx} = \big(\!\diag(\hat\vecr)\matF \kron \matI_B \big)\matC_{\quantize(\vecz)}\big(\matF^H\!\diag(\hat\vecr)^H \kron \matI_B \big)$. When computing the PSD for frequencies outside the range $[-f_\text{s}/2, f_\text{s}/2)$, we use that the PSD of the quantized signal is~periodic.~We note that our analytical results are in excellent agreement with the numerical~results.

\setlength{\textfloatsep}{10pt}
\begin{figure}[tp]
\centering
\includegraphics[width = .95\columnwidth]{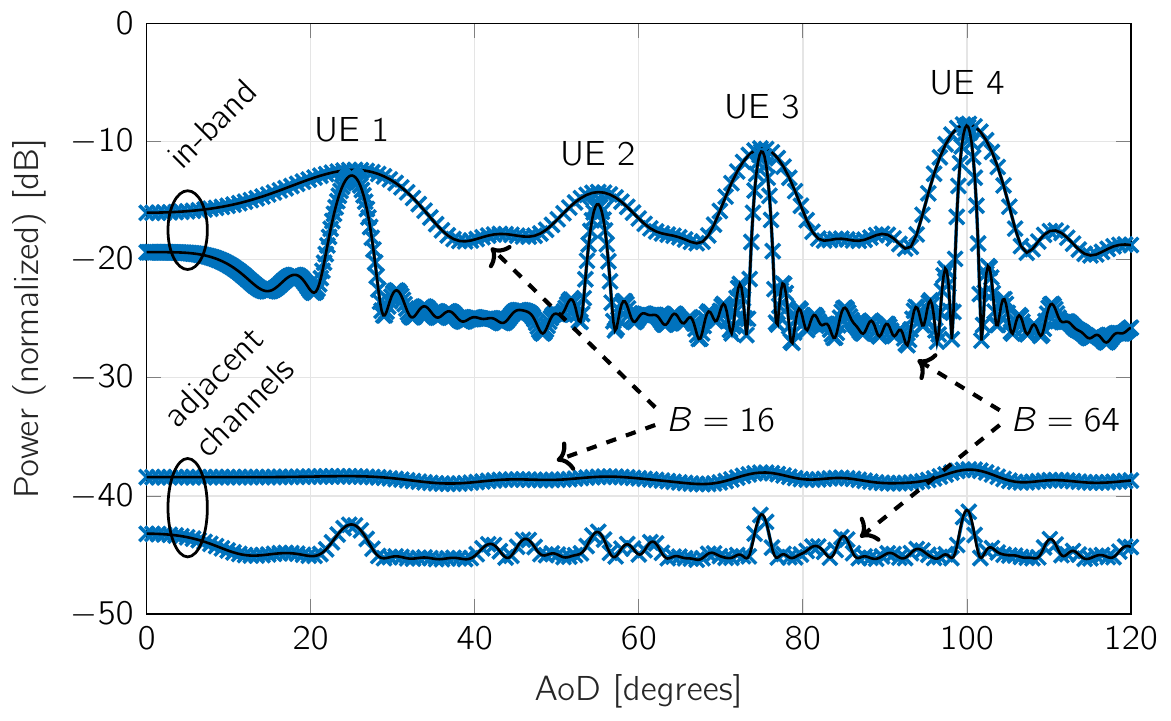}\label{fig:spatial}	
\vspace{0.07cm}
\caption{Radiation pattern for in-band and OOB (adjacent channels) emissions. 
The markers correspond to simulated values and the solid lines to analytical~results.
}
\label{fig:spatial}	
\end{figure}

In~\fref{fig:spatial},
we show the in-band and OOB radiation pattern (i.e., power radiated in different directions) for 1-bit DACs, when $\eta = 2$, $f_\text{cut} = 1.6875$~MHz, and $B\in\{16,64\}$
For a fair comparison between the $B=16$ and the $B=64$ cases, we have normalized the transmit power by $1/B$.
The in-band power radiated in the direction $\phi$ is~computed as $(S+1)^{-1}\sum_{b=1}^B\sum_{k \in \setS \union \{ 0 \}} \lefto[\matV(\phi)\matC_{\hat\vecx}\matV(\phi)^H\right]_{b + kB,b + kB}$
where $\matV(\phi)= \matI_N \kron \text{diag}(\vecv(\phi))$, and $\vecv(\phi)$ is the steering vector for the ULA in the direction $\phi$, i.e., the $b$th entry of $\vecv(\phi)$ is $\exp(-j\pi(b-1)\cos(\phi))$.
With a large number of antennas at the BS, we are able to steer beams in the direction of the~UEs. Indeed, we see that the in-band power has peaks at $25^\circ$, $55^\circ$, $75^\circ$, and~$100^\circ$ (where the UEs are). 
By increasing the number of antennas, the BS can create narrower beams, and less in-band power is radiated in unwanted directions (the side lobes are smaller).
We note that most power is radiated to the UE that is furthest away (UE~4) and that the least amount of power is directed towards the UE closest to the BS (UE~2). 

More interesting, perhaps, is to characterize the radiation pattern for the OOB emissions. In~\fref{fig:spatial}, we show the radiation pattern on  two \emph{adjacent channels}. The first adjacent channel is centered around $-5$~MHz, with transmission bandwidth~$f_\text{BW}$ (not including guard bands). The second adjacent channel is centered around $5$~MHz. 
The OOB power in the adjacent channels radiated in direction~$\phi$ is $({2S + 2})^{-1}\sum_{b=1}^B\sum_{k \in \setA_1 \union \setA_2} \lefto[\matV(\phi)\matC_{\hat\vecx}\matV(\phi)^H\right]_{b + kB, b + kB}$. Here, $\setA_1 = \{541, \dots, 841\}$ and $\setA_2 = \{183, \dots, 483\}$ are the set of subcarrier indices closest to the two adjacent channels, respectively. Interestingly, we note that the radiation pattern for the adjacent channels is reasonably flat,
although it has peaks in the direction of the four UEs  (especially for $B=64$). 
We also note that by increasing the number of BS antennas from $16$ to $64$, and by scaling down the transmit power accordingly, the OOB emissions radiated in the two adjacent channels is reduced in every direction.
This demonstrates that by increasing the number of antennas at the BS, OOB emissions due to low-resolution DACs can be significantly reduced.
Similar findings have been reported for PAs in~\cite{mollen16b}.

\subsection{Impact of Analog Filtering on the BER}

\setlength{\textfloatsep}{10pt}
\begin{figure}[tp]
	\centering
	\includegraphics[width=.95\columnwidth]{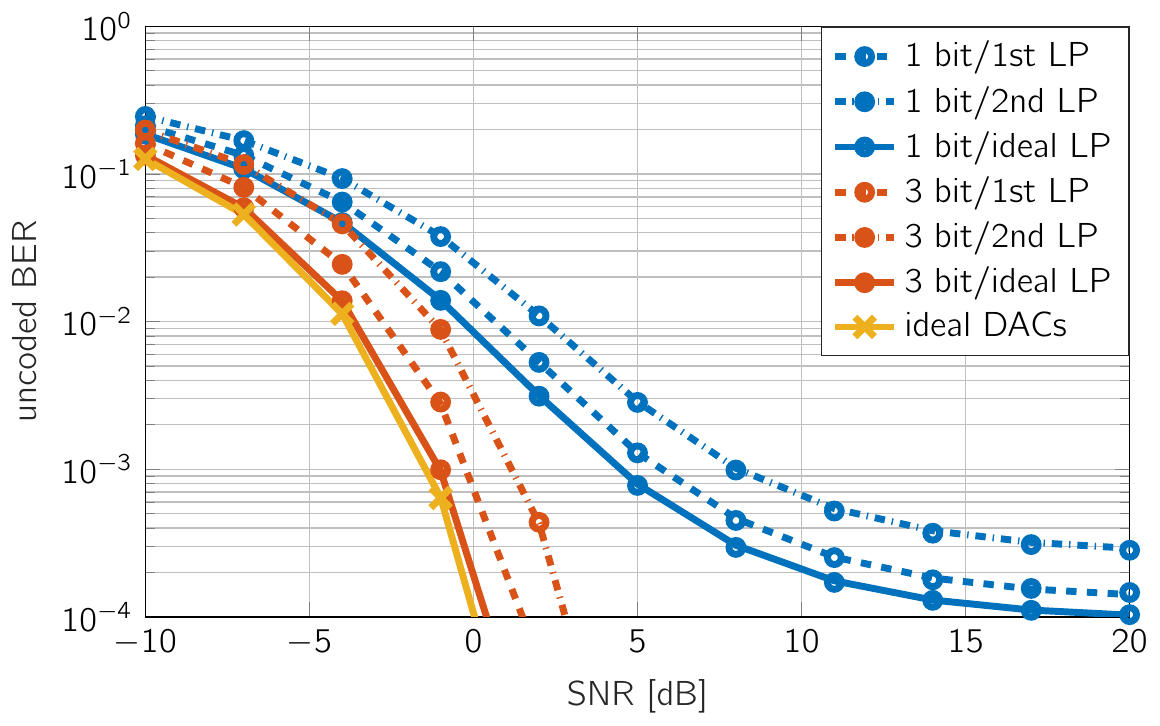}	
	\caption{Uncoded BER with QPSK and ZF (averaged over the UEs). The cut-off frequency of the 1st order filter is set to $f_\text{cut} = 2.25$~MHz, the cut-off frequency of the 2nd order Butterworth filter is set to $f_\text{cut} = 1.6875$~MHz. The markers correspond to simulated values and the lines to analytical results.}
	\label{fig:ber}
\end{figure}

In \fref{fig:ber}, we investigate the impact of coarse quantization and analog filtering on the BER in the massive MU-MIMO-OFDM downlink.
%
Specifically, we show the uncoded BER (averaged over the UEs) with QPSK and ZF as a function of the signal-to-noise ratio~(SNR), defined as $\textit{SNR} = 1/N_0$. 
We consider 1-bit DACs and 3-bit DACs and  different LP filters. 
As a reference, we also show the BER when the reconstruction stage is an ideal LP filter (i.e., the case studied in~\cite{jacobsson17c, jacobsson17e}) and with ideal DACs (i.e., infinite resolution and ideal LP filter). 
Numerical simulations are compared to our analytical results. 
Specifically, for a given channel realization, the uncoded BER with QPSK is computed using the SINDR in~\eqref{eq:sindr} as~follows:
\begin{IEEEeqnarray}{rCl}
	\textit{BER} &=& 1 - \frac{1}{US} \sum_{u=1}^U \sum_{k \in \setS} \Phi \lefto( \sqrt{\textit{SINDR}_{u,k}\big( \widehat\matH\big)}\,\right).
\end{IEEEeqnarray}
Here, $\Phi(x) = \frac{1}{\sqrt{2\pi}}\int_{-\infty}^x e^{-u^2/2} \text{d}t$.

Using sharper filters and reducing the cut-off frequency to attenuate OOB emissions results---as expected---in a BER degradation. 
We note, however, that this degradation is limited.
For example, when $\eta = 1$ and $f_\text{cut} = 2.25$~MHz, the loss is about $ 0.7$\,dB for 3-bit DACs at a target uncoded BER of $10^{-3}$.

\subsection{Tradeoffs between SINDR, ACLR, and PAR} \label{sec:numerical_tradeoff}

\setlength{\textfloatsep}{10pt}
\begin{figure}[!t]
	\centering
	\subfloat[Average ACLR vs average SINDR. Here, the cut-off frequency \emph{increases} from left to right.]{\includegraphics[width=.475\columnwidth]{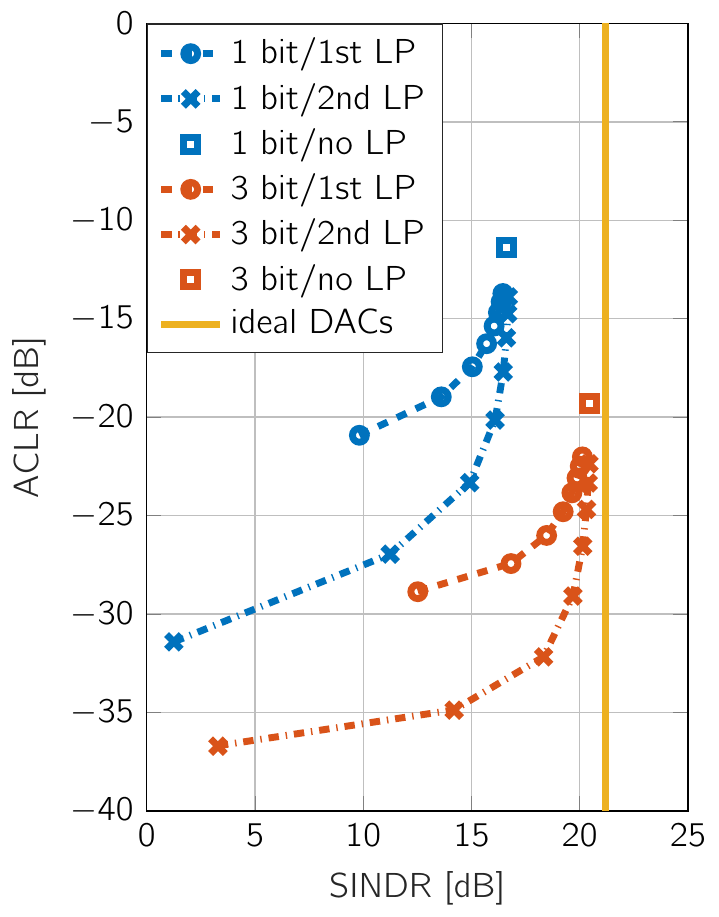}\label{fig:aclr}} \quad
	\subfloat[Average ACLR vs average PAR. Here, the cut-off frequency \emph{decreases} from left to right.]{\includegraphics[width=.475\columnwidth]{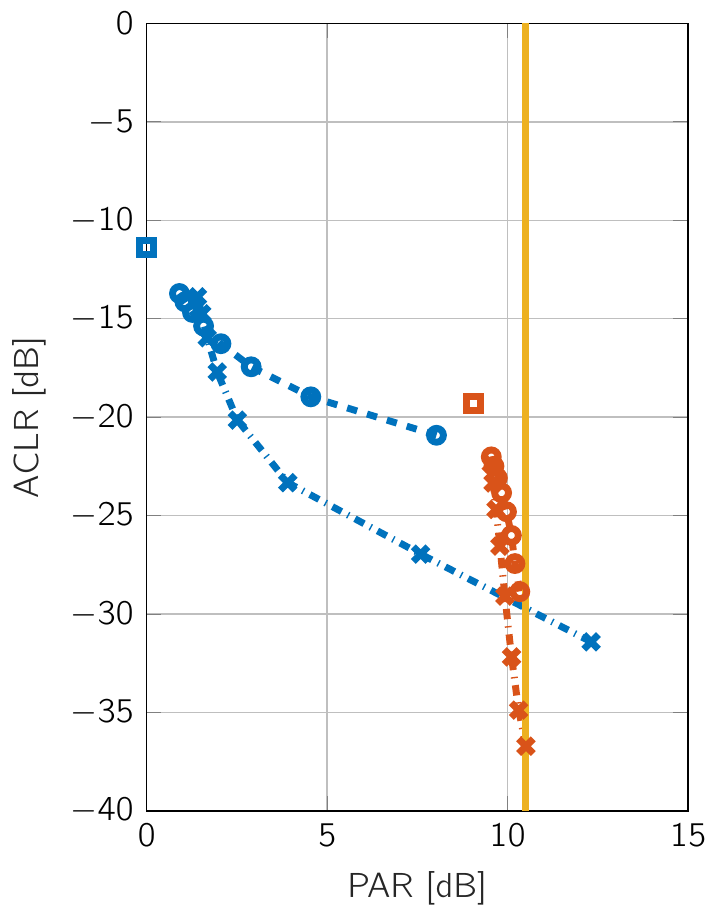}\label{fig:par}}
	\vspace{0.1cm}
	\caption{Tradeoffs between ACLR, SINDR, and PAR. The cut-off frequency of the LP filter is swept from $562.5$~kHz to $4.5$~MHz in increments of $562.5$~kHz. The markers correspond to simulated ACLR values and the lines to analytical~results.}
	\label{fig:trade-off}
\end{figure}

%
In~\fref{fig:aclr}, we plot the ACLR (analytical and simulated) as a function of the SINDR, computed using~\eqref{eq:sindr}, for 1-bit and 3-bit DACs, and for  $\eta = \{1,2\}$. 
The cut-off frequency of the LP filter is swept from $562.5$~kHz to $4.5$~MHZ in increments of $562.5$~kHz and the SNR is set to $\textit{SNR} = 10$~dB.
The ACLR is computed as $\textit{ACLR} = \frac{1}{B} \sum_{b=1}^B \textit{ACLR}_b$, where $\textit{ACLR}_b$ is the ACLR on the $b$th BS antenna:
\begin{IEEEeqnarray}{rCl}
\textit{ACLR}_b
&=& \frac{\text{max}\lefto\{\int_{\setF_1} \Ex{}{\lvert \hat{x}_b(f) \rvert^2}\text{d}f,\int_{\setF_2} \Ex{}{\lvert \hat{x}_b(f) \rvert^2}\text{d}f\right\}}{\int_{-\Delta{f}S/2}^{\Delta{f}S/2} \Ex{}{\lvert \hat{x}_b(f) \rvert^2}\text{d}f} \IEEEeqnarraynumspace \\
&\approx&\frac{\text{max}\lefto\{\sum_{\setA_1} P_{\hat{x}_{b,k}}, \sum_{\setA_2} P_{\hat{x}_{b,k}}\right\}}{\sum_{\setS \union \{ 0 \}} P_{\hat{x}_{b,k}}}. \IEEEeqnarraynumspace 
\end{IEEEeqnarray}
Here, $\hat{x}_b(f) = \int_{-\infty}^\infty x_b(t)e^{-j2\pi f t} \text{d}f$. Furthermore, $\setF_1 = (-7.25, -2.75)$~MHz and $\setF_2 = (2.75, 7.25)$~MHz are the first and the second adjacent channel, respectively.
We see from~\fref{fig:aclr} that reducing the cut-off frequency improves the ACLR but decreases the SINDR, i.e., there is a tradeoff between ACLR and SINDR. 
For example, for the 3-bit DACs case, using $\eta = 2$ and $f_\text{cut} = 1.125$~MHz results in an ACLR improvement of over $15$~dB compared to the unfiltered case (``3 bit/no LP''), but also in a loss of about $6$~dB in SINDR.

In~\fref{fig:par}, we illustrate the tradeoff between ACLR (analytical and simulated) and PAR.
Here, we have again swept the cut-off frequency of the LP filter from $562.5$~kHz to $4.5$~MHz in increments of $562.5$~kHz. 
The PAR is computed as $\textit{PAR} = \frac{1}{B}\sum_{b=1}^B \textit{PAR}_b$, where $\textit{PAR}_b$ is the PAR on the $b$th BS antenna:
\begin{IEEEeqnarray}{rCl}
	\textit{PAR}_b = \frac{2N \opnorm{\lefto[x_{b,0},\dots, x_{b,N-1}\right]^T}_{\widetilde\infty}^2}{\opnorm{\lefto[x_{b,0},\dots, x_{b,N-1}\right]^T}_2^2}.
\end{IEEEeqnarray}
A low  PAR value is desirable as it enables more efficient PA designs~\cite{mollen16e}.
Note that with 1-bit DACs and when the reconstruction stage is a ZOH filter only (``1 bit/no LP'') the resulting waveform has $0$~dB PAR.
By introducing analog filters to lower the ACLR, we do not only lose in terms of SINDR (see~\fref{fig:aclr}) but also in terms of PAR. 
For example, with 1-bit DACs and LP filtering, the transmitted waveform has no longer $0$~dB PAR. 
However,  the resulting waveform  has still a significantly lower PAR compared to the ideal DACs case. 
This implies that 1-bit DACs enable power-efficient PA designs.

\section{Conclusions}

We have investigated the OOB emissions caused by low-resolution DACs in massive MU-MIMO-OFDM, and shown how practical analog filters (e.g., a ZOH filter followed by a second-order Butterworth filter) can be used to mitigate these OOB emissions.
Such filters create in-band signal distortion, which can be reduced effectively using simple DPD schemes.

We have also analyzed the tradeoffs between ACLR, SINDR, and PAR. By using higher-order filters and by decreasing the cut-off frequency, we gain in terms of ACLR (OOB emissions are attenuated) but we lose in terms of SINDR (the BER increases) and PAR. 
By carefully tuning the analog filter parameters, one can achieve satisfactory OOB performance with acceptable  BER and PAR degradation. Finally, in agreement with~\cite{mollen16b}, we have shown that the transmit-power reduction enabled by the use of large antenna arrays at the BS-side yields also a significant reduction in OOB~emissions. 
%

%
%
%
%


\begin{spacing}{0.897}
\bibliographystyle{IEEEtran}
\bibliography{IEEEabrv,confs-jrnls,publishers,svenbib}	
\end{spacing}

\end{document}